\documentclass[aps,prl,preprint,showpacs,groupedaddress]{revtex4-1}
\usepackage{amssymb}
\usepackage{graphicx}
\usepackage{amsmath}

\usepackage{color}

\begin{document}

\setcounter{page}{1}

\title{Opposing effects of stacking faults and antisite domain boundaries on the conduction band edge in kesterite quaternary semiconductors}

\author{Ji-Sang Park}
\affiliation{Thomas Young Centre and Department of Materials, Imperial College London, Exhibition Road, London SW7 2AZ, UK}
\email[]{ji-sang.park@imperial.ac.uk}

\author{Sunghyun Kim}
\affiliation{Thomas Young Centre and Department of Materials, Imperial College London, Exhibition Road, London SW7 2AZ, UK}

\author{Aron Walsh}
\affiliation{Thomas Young Centre and Department of Materials, Imperial College London, Exhibition Road, London SW7 2AZ, UK}
\affiliation{Department of Materials Science and Engineering, Yonsei University, Seoul 03722, Korea}

\bibliographystyle{apsrev4-1}

\date{\today}
\begin{abstract}

We investigated stability and the electronic structure of extended defects including anti-site domain boundaries and stacking faults in the kesterite-structured semiconductors, Cu$_2$ZnSnS$_4$ (CZTS) and Cu$_2$ZnSnSe$_4$ (CZTSe).
Our hybrid density functional theory calculations show that stacking faults in CZTS and CZTSe induce a higher conduction band edge than the bulk counterparts, and thus the stacking faults act as electron barriers.
Antisite domain boundaries, however, accumulate electrons as the conduction band edge is reduced in energy, having an opposite role.
An Ising model was constructed to account for the stability of stacking faults, which shows the nearest neighbour interaction is stronger in the case of the selenide.

\end{abstract}

\maketitle

Thin-film photovoltaic devices based on Cu$_2$ZnSn(S,Se)$_4$ (CZTSSe) absorber layers have attracted growing attention  \cite{polizzotti2013state,walsh2012kesterite,wallace2017steady} as the materials are composed of Earth-abundant elements \cite{schmalensee2015future}, which are not categorised as Critical Raw Materials (CRM) by EU \cite{eu2014report}.
The system has a tuneable direct band gap of 1.0$\sim$1.5 eV \cite{chen2009crystal}, which is ideal for single junction solar cell applications \cite{shockley1961detailed}.
The certified solar conversion efficiency of 12.6\% was achieved by an IBM group in 2013 \cite{AENM201301465},
and more recently, another group at DGIST achieved an efficiency of 12.3 \% in 2016 by using a band-gap-graded absorber layer \cite{yang2016band}.

Since current thin-film technologies mostly rely on polycrystalline materials, physical properties of extended defects, especially grain boundaries (GBs) have been investigated to understand their effects on the device efficiency \cite{wang2011structural,li2012investigating,mendis2012role,kim2014surface,yin2014engineering,gershon2015role,liu2017sodium}.
Other extended defects like stacking faults (SFs) and antisite domain boundaries (ADBs) have been less documented as compared to the GBs,
but since SFs in CdTe act as electron barriers and reduce the efficiency \cite{Yan2001,Abbas2013,Sun2013,Yoo2014a}, SFs in CZTS should be investigated.
There is also growing evidence that the materials have extended defects \cite{song2015epitaxial,kattan2015crystal,kattan2016observation}.
Formation of SFs was found in CZTS grown on single crystal Si (111) wafers \cite{song2015epitaxial} and CZTS nanoparticles \cite{kattan2015crystal}.
Another recent experimental study has shown that ADBs are formed abundantly in CZTS nanocrystals \cite{kattan2016observation}, possibly due to the low formation energy of antisite defect complexes in multi-component semiconductors \cite{walsh2012kesterite}.
A density functional theory (DFT) calculation also shows that pre-existing defect complexes can lower the energy cost to form another defect complexes in close configuration \cite{huang2013band}, providing a hint that point defects can be gathered and form a spatially extended defect.

In this study, we investigate stability and the electronic structure of extended defects including SFs, ADBs, and the $\Sigma$3 (112) GB. 
We constructed an Ising model to account for the stability of SFs and examined an effect of broken symmetry at the boundary on the electronic structure. 
Our results show that the formation energy of SFs is small, while it is well explained by the Ising model. 
Change of the stacking orders raises the conduction band minimum (CBM) and thus the SFs generally act as electron barrier.
On the other hand, the ADB with $\frac 12[110]$ fault displacement induces several ten meV lower conduction band edge than the bulk counterpart, indicating that the defect could be a place where electrons are temporarily trapped.

We performed first-principles density functional theory (DFT) calculations to investigate physical properties of the extended defects.
The hybrid functional proposed by Heyd, Scuseria, and Ernzerhof \cite{HSE} as implemented in the VASP code was used \cite{PhysRevB.54.11169}.
The projector-augmented wave (PAW) pseudo-potentials were used to describe the valence and core electron interactions \cite{PAW}.
The screening parameter of 0.2 {\AA}$^{-1}$ and the exchange parameter of $\alpha$ = 0.25 were used. The cutoff energy for the plane-wave basis was set to 400 eV.
The lattice parameters and the internal coordinates were fully relaxed until the residual force becomes smaller than 0.03 eV {\AA}$^{-1}$.
For Brillouin zone (BZ) integration, the smallest spacing between \textit{k}-points was set to $\simeq$0.05 {\AA}$^{-1}$.

The atomic structure of SFs and the $\Sigma$3 (112) GB are shown in Figure 1.
We note that each layer in the supercells has two Cu, one Zn, and one Sn atoms. 
Therefore, the position of the cations in an adjacent layer is determined when the Octet rule is preserved.
Among various stacking faults, 9R, intrinsic stacking fault (ISF) and extrinsic stacking fault (eSF) were considered. 
The SFs has stacking sequences of ($\cdots$ABC/BCA/CAB$\cdots$), ($\cdots$ABC/BC/ABC$\cdots$), and ($\cdots$ABC/ABAC/ABC$\cdots$), respectively, as depicted in Figure 1, thus one can generate a supercell having a SF.
On the other hand, a supercell having a $\Sigma$3 (112) GB contains two GBs because the $\Sigma$3 (112) GB has a layer with inversion symmetry at the middle of the cell ($\cdots$ABA$\cdots$).
Another $\Sigma$3 (112) GB (e.g. $\cdots$ACA$\cdots$) is needed to restore the sequence order. Otherwise, a slab geometry should be pursued.

On the other hand, the ADBs can be represented by the accumulation of cation antisites in planes. 
Thus the Octet rule may or may not be satisfied at an ADB, depending on the fault displacement. 
For instance, Kattan \textit{et al.} reported atomic structures of three ADBs, one satisfying the Octet rule and the others not satisfying the rule \cite{kattan2016observation}.
Among them, we generated the atomic structures of ADBs with fault displacement of $\frac 12[110]$ or $-\frac12[201]$, which are shown in Figure 1e and Figure 1f. 
Despite that the former is called an ADB, 
the Octet rule is not broken as its structure can be generated from kesterite by shifting a group of layers by ($a$/2,$a$/2,0) where $a$ is the lattice constant along $x$ and $y$ directions.
As a result, narrow planes with Cu atoms are formed at the boundary. 
Such planes are also formed in primitive-mixed CuAu phases (PMCA), which another polytype of CZTS \cite{chen2009crystal}.
Generally speaking, such faults in this category of materials results in higher formation energy and lower band gap, predicted by a previous first-principles calculation \cite{park2015ordering}.
The Octet rule is broken at the other ADBs, and thus some S or Se atoms are bonded to 2 Sn atoms (The coordination in bulk kesterite is 1 Sn, 1 Zn, and 2 Cu).

To investigate the thermodynamic stability of the SFs and the $\Sigma$3 (112) GB, we constructed an Ising model following an approach which was used to understand polytypes of SiC \cite{rutter1997energetics}.
Our Ising model for a supercell with $N$ layers is given by
\begin{equation}
    E_{tot} = J_0 N  + \sum^M_{n=1} \sum^{N}_{i=1} J_n \sigma_i \sigma_{i+n},
    \end{equation}
where $E_{tot}$ is the total energy of a given supercell. The energy of a single layer is given by $J_0$, and $J_n$ represents the interaction energy between the $n$th nearest neighbour layers ($n$ = 1, 2, $\cdots$, M). 
An $i$th layer can have either spin up ($\sigma=1$) or spin down ($\sigma=-1$), which is determined by comparison to the next layer ($i+1$th layer).
If two adjoining layers have AB, BC or CA stacking order, then the first layer has spin up. The two layers do not have the same letters (i.e. AA, BB and CC) in this study, and spin down is assigned to the $i$th spin in remaining cases.
Since we use periodic boundary conditions, $\sigma_1 = \sigma_{N+1}$, the total energy of bulk is equivalent to $J_0$+$J_1$+$J_2$+$J_3$ per layer when $M$ is equal to 3.

The fitted parameters for SFs in CZTS are J$_1$ = -20 meV/nm$^2$, J$_2$ = 0 meV/nm$^2$ and J$_3$ = 1 meV/nm$^2$. 
On the other hand, those for SFs in CZTSe are J$_1$ = -31 meV/nm$^2$, J$_2$ = 6 meV/nm$^2$ and J$_3$ = -3 meV/nm$^2$.
We don't report $J_0$ because the absolute value of $J_0$ doesn't have the physical meaning in our DFT calculation and the relative energy of SFs can be calculated without knowing $J_0$.
The strongest interaction parameters $J_1$ is significantly larger in CZTSe, indicating that SFs are less likely formed in CZTSe.
This tendency is largely depicted in high formation energy of the wurtzite phase (2H) in CZTSe than CZTS,
and also consistent with the anion rule that the zinc-blende phase becomes more favourable than the wurtzite phase as the anion size increases \cite{PhysRevB.46.10086}.
Another difference between CZTS and CZTSe is smaller values of $J_1$ and $J_2$, which results in the similar formation energy of 2H and 4H in CZTS.

Using the raw data obtained from DFT calculations and the Ising model, we calculated the formation energy of the extended defects. The formation energy of a SF, $E_{f}(\mathrm{SF})$, is given by
\begin{equation}
    E_{f}(\mathrm{SF}) = \frac {E_{tot}(\mathrm{SF})}{A} - \frac{NE_{tot}(\mathrm{0})}{A},
    \end{equation}
where $N$ is the number of layers in a supercell and $A$ is a unit area of the SF. $E_{tot}(\mathrm{0})$ is total energy of bulk per unit cell (8 atoms).
The formation energy obtained from DFT calculations ($E_{f,\mathrm{DFT}}$) and that obtained from the Ising model ($E_{f,\mathrm{Ising}}$) are summarised in Table 1.
The difference between them is small enough to conclude that the formation energy of SFs is well explained by the Ising model.
Stability of the two extreme cases, AB and ABCB, is well explained by the Ising model even though the two configurations were not considered to obtain the parameters $J_n$. 
The calculated $E_f$ values of SFs and the GB are small, which is also consistent with other studies reporting the formation of SFs in other materials like Si and CdTe \cite{Chou1985,Kackell1998,Yan2001,Yoo2014a,park2015stability}.

To examine how the band edges are affected by the extended defects, we obtained averaged local potential given as
\begin{equation}
\overline{V(z_0)} = \frac {{\int_{z_0-\tau/2}^{z_0+\tau/2} \int \int  V(x,y,z) dxdydz }} { \tau  \int\int dxdy},
\end{equation}
where $\tau$ is the interlayer distance. 
Band edges of pure CZTS or CZTSe were estimated using $\overline{V(z_0)}$ in a bulk-like region as a reference. 
The valence band offset (VBO) and the conduction band offset (CBO) between polytypes and the bulk counterparts are summarised in Table 1.
There is no bulk-like region in (AB) and (ABCB), and thus the band offsets are not calculated.
In both CZTS and CZTSe, the VBO is smaller than the room temperature energy. 
Therefore, we expect that it will not significantly affect the hole transport. 
On the other hand, the conduction band of the material with SFs are higher than that of the bulk counterpart, which is comparable to the room temperature thermal energy. 
This result clearly indicates that SFs act as electron barrier, making electron extraction difficult.
It is generally accepted that SFs in zinc-blende structure (ABC) can be understood as a thin wurtzite layer (AB) surrounded by zinc-blende grains, and that results in electron barrier because of the type-II band offset between wurtzite and zinc-blende \cite{murayama1994chemical,yan2007understanding}, which is also found in the multi-component semiconductors.
Our result is also consistent with the higher band gap of wurtzite-kesterite CZTS than kesterite CZTS and group theory analysis \cite{chen2010wurtzite}.

The stability of ADBs suggested by an experimental study \cite{kattan2016observation} were also investigated.
It is worth emphasising that the suggested ADBs are formed only in multi-cation semiconductors as the ADBs are represented by cation disorder and thus don't have anion-anion or cation-cation bonds.
Supercell containing a $\frac 12[110]$ ADB is stoichiometric, therefore the formation energy of the defect is simply calculated as defined above. 
$E_{f}$ is 0.23 eV/nm$^2$ and 0.49 eV/nm$^2$ in CZTS and CZTSe, respectively. 
Higher energy is required to form the ADB in CZTSe as compared to that in CZTS, indicating that the ADB is also less likely formed in CZTSe.
On the electronic structure, the VBO between the $\frac 12[110]$ ADB and bulk calculated using the potential alignments are negligible in both CZTS and CZTSe ($<$ 3 meV).
The CBO, on the other hand, is -70 meV and -77 meV, respectively, indicating that the ADBs can easily trap electrons, which is \textit{opposite} to the SFs.

This opposite effect of the ADB on the conduction band is consistent with a previous DFT calculation with symmetry analysis \cite{park2015ordering}.
The ADB, a polytype with infinite length, has the similar atomic structure to PMCA in a local sense as both have Cu layers, while the $\frac 12[110]$ faults increase the formation energy of polytype as shown in a previous study \cite{park2015ordering}.
The electronic band gap of the polytype is negative-linearly correlated with the formation energy of polytypes \cite{park2015ordering}, and thus the ADB should lower the band gap.
Moreover, as it has been shown that the kesterite, stannite, and PMCA Cu$_2$ZnGeS$_4$ have similar valence band edge position, and change of the band gap is mainly explained by change of the conduction band \cite{PhysRevB.79.165211}.
Since the ADB can be regarded as high energy polytype in a local sense, the $\frac 12[110]$ ADB is expected to have lower conduction band edge than bulk, which is found in our calculation.
We note that the band offset due to the ADB is similar to the band gap fluctuations in real samples (0.05-0.15 eV) \cite{bishop2017modification,gokmen2013band}, indicating that the V$\mathrm{_{OC}}$ deficit can be at least partly explained by the formation of the ADBs.

Experimental evidence of faults in the layer with Cu and Sn has been provided \cite{kattan2016observation}, however, Cu and Zn are difficult to distinguish by transmission electron microscope (TEM). 
Due to the larger chemical (size and charge) mismatch, the layer with Cu and Sn should be more rigid than the layer with Cu and Zn.
But the effect of Cu-Zn disorder on the electronic structure should be less than 0.04 eV according to the DFT calculation \cite{PhysRevB.79.165211}.
We also examined whether Zn$_{\mathrm{Cu}}$+Cu$_{\mathrm{Zn}}$ at the $\frac 12[110]$ ADB affects the conclusion, but it changes the band edge position only marginally ($\simeq$ 3 meV).

The supercell model for the ADB with $-\frac12[201]$ fault (Figure 1f) contains more Zn and Sn atoms as compared to bulk, and thus the boundary can be understood as segregation of Sn$_{\mathrm{Cu}}$ and a Zn$_{\mathrm{Cu}}$ defects.
Relaxation of internal coordinates for the supercell model using HSE06 functional is too computationally demanding, therefore, we relaxed the structure using the SCAN \cite{SCAN} functional applying on-site Coulomb potential of 6 eV on Cu \textit{d} and Zn \textit{d}.
Self-consistent field (SCF) calculations using HSE06 functional were subsequently performed to analyse the electronic structure.

The calculated projected density of states (PDOS) of the ADB in CZTS and CZTSe with $-\frac12[201]$ fault are shown in Figure 2.   
We find that Sn atoms at the boundary introduce gap states, which are mainly composed of Sn s and S (Se) p anti-bonding character.
Our supercell is not large enough to reproduce the bulk band gap, however, we expect that the band gap is widened in the supercell because of the raised conduction band resulting from the quantum confinement effect.
Consistent with our expectation, the PDOS of Cu in bulk CZTS fits well with that in the supercell.
In CZTSe, peak positions are slightly shifted ($\simeq$ 0.1 eV), but doesn't affect the conclusion that the ADB introduces the gap states
as the lowest defect state is higher than the valence band maximum by 0.52 eV, close to the middle of the band gap.
The gap states are delocalized in the boundary, indicating that charge carriers trapped by the state will conduct in the boundary.

A post-deposition annealing treatment (e.g. using CdCl$_2$) is necessary to improve the CdTe solar cells \cite{Major2014}.
There are many competing explanations on the beneficial effect of the treatment \cite{Dharmadasa2014}, and recent studies also show that one effect of the treatment is the removal of the SFs \cite{Abbas2013,Yoo2014a}.
A SF in CdTe can be regarded as a buried wurtzite phase which has higher conduction band than the zinc-blende CdTe, and thus it is expected to act as an electron barrier \cite{wei2000}.
It was recently found in an experimental study that the conductivity along the direction normal to the SFs is suppressed because of the band offset \cite{Sun2013}. 
Since the SFs in the kesterite-structured materials also act as electron barriers, a similar annealing process should be pursued to remove the extended defects from the absorber layer.

Finally, we point out that the electrical property of the SFs can be qualitatively estimated as discussed above by comparing the band edges of polymorphs, and this working principle is not limited to zinc-blende and zinc-blende derived structures. For instance, previous DFT calculations show that wurtzite ZnO and III-nitrides have higher conduction band than their zinc-blende counterparts, and thus the SFs lower the conduction bands and act as electron sinks, not barriers as in CZTS \cite{yan2004energetics, stampfl1998energetics}.

In summary, we investigated the thermodynamic stability and the electronic structure of extended defects in the multi-cation semiconductors, CZTS and CZTSe.
Formation energy of extended defects in CZTS and CZTSe were calculated by performing hybrid density functional theory calculations. 
Since less energy is required to form SFs than the ADBs, SFs are more likely formed in the multi-cation semiconductors.
An Ising model was successfully constructed to account for their stability, and the interaction between two adjacent layers is fitted to be stronger than the other interactions between layers.
The SFs and the ADBs satisfying the Octet rule introduce higher and lower conduction band than the bulk region, acting as electron barrier and sink, respectively. 
The ADB not satisfying the Octet rule, on the other hand, introduces deep gap states.
Compared to the electron transport, the hole transport is less affected by the extended defects.
Our computational results indicate that extended defects slightly favoured in CZTS as compared to CZTSe, potentially results in larger variation of the conduction band edge.
Annealing procedures used for other technologies (e.g. CdCl$_2$ for CdTe) could be applied to the kesterite solar cells.

\begin{acknowledgments}
This project has received funding from the European H2020 Framework Programme for research, technological development and demonstration under grant agreement no. 720907. See http://www.starcell.eu. 
AW is supported by a Royal Society University Research Fellowship. 
Via our membership of the UK's HPC Materials Chemistry Consortium, which is funded by EPSRC (EP/L000202), this work used the ARCHER UK National Supercomputing Service (http://www.archer.ac.uk). We are grateful to the UK Materials and Molecular Modelling Hub for computational resources, which is partially funded by EPSRC (EP/P020194/1). 
\end{acknowledgments}

\bibliography{czts}

\clearpage
\begin{table}
\begin{center}
\caption{Physical properties of CZTS with different stacking orders. 
$E_{f,\mathrm{DFT}}$ and $E_{f,\mathrm{Ising}}$ are the formation energy obtained by DFT calculations and that estimated by the Ising model, respectively, which are defined in eq. (2). The dimension of the formation energies is eV/nm$^2$. Values in parentheses are those of CZTSe. The valence and conduction band offsets with respect to the bulk material are labelled as VBO and CBO, respectively.}
\begin{tabular}{llccccc}
\hline \hline 
	Stacking     &
	$E_{f,\mathrm{DFT}}$ & 
	$E_{f,\mathrm{Ising}}$ & 
	VBO (meV) & 
	CBO (meV) \\ \hline \hline

	ISF & 0.14 (0.18)
	& 0.15 (0.20)
	&
        6 (9) &
	28 (29) \\ \hline

	eSF & 0.14 (0.16)
	& 0.15 (0.17)
	&
	6 (5) &
	28 (25) \\ \hline

	9R & 0.31 (0.37)
	& 0.31 (0.37)
	&
	-2 (11) &
	32 (48) \\ \hline

	$\Sigma$3 (112) & 0.07 (0.10)
	& 0.07 (0.10)
	&
	-15 (-5) &
	16 (18) \\ \hline

	2H (AB) & 0.17 (0.25)
	& 0.15 (0.24)
	\\ \hline

        4H (ABCB) & 0.15 (0.17)
	& 0.15 (0.15)
	\\ \hline
\hline 
\end{tabular}

\end{center}
\end{table}

\clearpage

\begin{figure}
\includegraphics[width=\columnwidth]{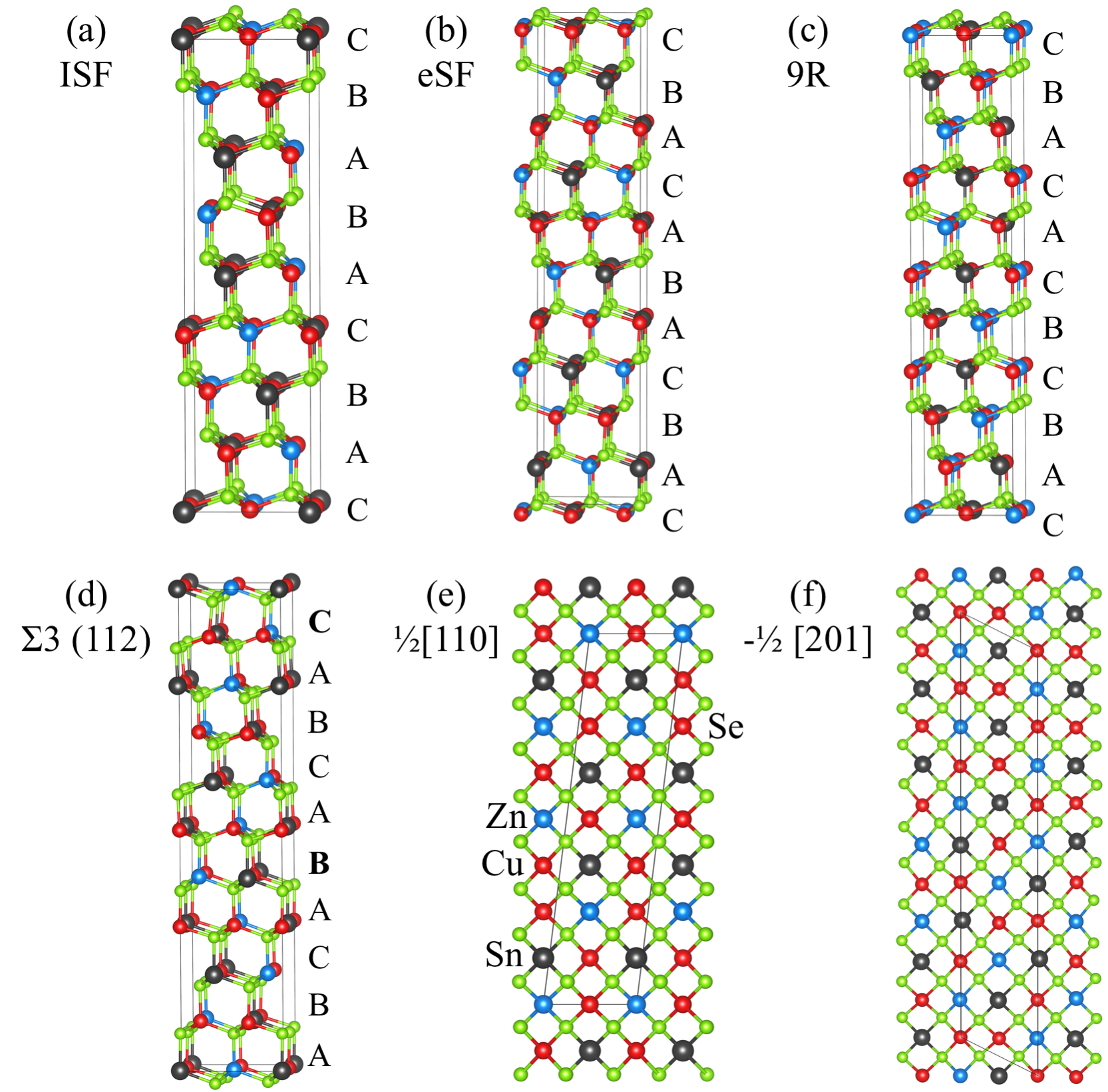}
\caption{\label{fig:1} (a-c) Atomic structure of intrinsic stacking faults, extrinsic stacking fault, and 9R. (d) Atomic structure of the $\Sigma$3 (112) grain boundary. (e,f) Atomic structure of antisite domain boundaries with the fault displacement of $\frac{1}{2} [110]$ or $-\frac{1}{2} [201]$. Solid lines represent the boundaries of the supercells.}
\end{figure}

\begin{figure}
\includegraphics[width=\columnwidth,scale=0.9]{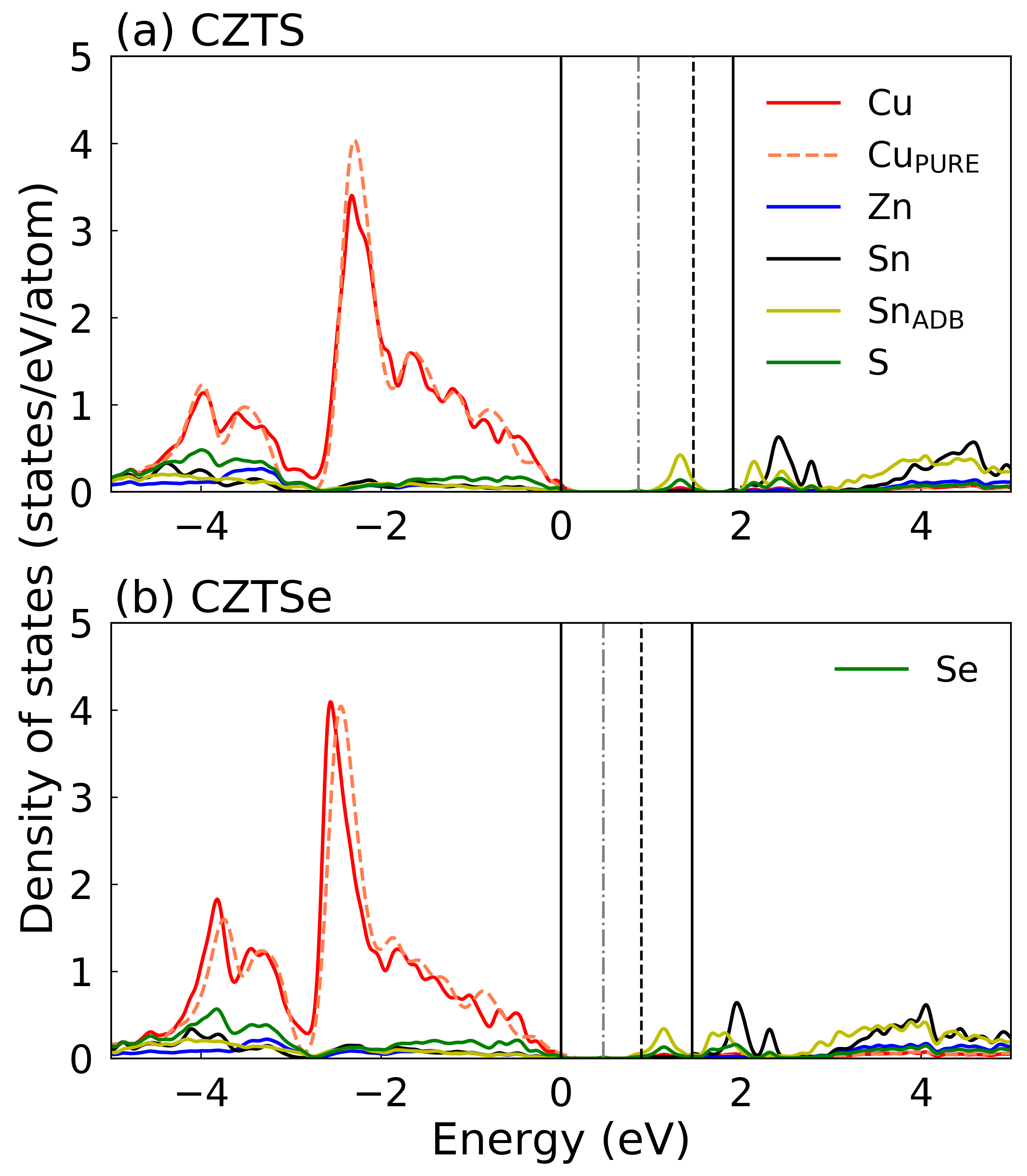}
\caption{\label{fig:2} Project density of states (PDOS) of the ADB with $-\frac12[201]$ fault. 
PDOS of Sn atoms in the boundary and bulk-like region are labelled as Sn$_{\mathrm{ADB}}$ and Sn, respectively. 
Vertical lines represent the band edges in each supercell.
Dashed lines are the estimated conduction band minimum using the band gap and the valence band maximum in the supercell. Dash dot lines represent the lowest defects states of the supercell. 
PDOS of Cu in bulk CZTS and CZTSe (Cu$_{\mathrm{PURE}}$) is shown for comparison.
}
\end{figure}
\end{document}